\documentclass[12pt]{article}
\usepackage{amssymb}
\usepackage{mathrsfs}
\usepackage{bm}
\usepackage{graphicx}
\usepackage{subfigure}
\usepackage{hyperref}
\usepackage{cite}
\usepackage[utf8]{inputenc}
\begin{document}
\begin{center}
\begin{Large}
{\bf Supersymmetry and deterministic chaos}
\end{Large}
\vskip1truecm
Stam Nicolis\footnote{E-Mail: stam.nicolis@idpoisson.fr; stam.nicolis@lmpt.univ-tours.fr}

{\sl Institut Denis Poisson, 
Universit\'e de Tours, Universit\'e d'Orl\'eans, CNRS\\
Parc Grandmont, Tours 37200, France}

\end{center}

\vskip1truecm

\begin{abstract}
We show that the fluctuations of the periodic orbits of deterministically chaotic systems can be captured by supersymmetry, in the sense that they are repackaged 
in the contribution of the absolute value of the determinant of the noise fields, defined by the equations of motion.  
\end{abstract}
\section{Introduction}\label{intro}
It is possible to interpret a set of first order ordinary differential equations, in the variables $x_I(t)$, 
\begin{equation}
\label{ODEs}
\frac{dx_I}{dt}=F_I(\bm{x})
\end{equation}
where $I=1,2,\ldots,d$ as describing the motion of a particle, as it probes $d-$dimensional space, where $x_I(t)$ is its trajectory.

The reason is that these equations can be understood as describing the minimization of the functional
\begin{equation}
\label{action}
S=\int\,dt\,\sum_{I=1}^d\,\frac{1}{2}\left(\dot{x}_I-F_I(\bm{x})\right)^2
\end{equation}
where $S$ is the Euclidian action. 

Their solutions saturate the bound $S\geq 0$.

If $F_I$ can be expressed as the gradient of a scalar function, i.e. $F_I = -\partial_I W$, then, upon expanding the integrand, we realize that it takes the form
\begin{equation}
\label{potentialmot}
\mathscr{L}=\frac{1}{2}\sum_{I=1}^d\,[\dot{x}_I]^2+\frac{1}{2}\sum_{I=1}^d\,\left(\frac{\partial W}{\partial x_I}\right)^2+\sum_{I=1}^d\,\dot{x}_I\frac{\partial W}{\partial x_I}
\end{equation}
The middle term can be identified with the scalar potential, $V(\bm{x})$, while the last term is a total derivative. 

The canonical partition function is well--defined, upon imposing the periodic boundary conditions, that eliminate the total derivative, if it is true that $V(\bm{x})$ confines at infinity (that it's bounded from below is obvious) and the fluctuations can be consistently described by the property that the ``noise fields'', 
\begin{equation}
\label{noisefields}
\eta_I\equiv \dot{x}_I - F_I
\end{equation}
are Gaussian fields, with ultra--local 2--point function, $\langle\eta_I(t)\eta_J(t')\rangle\propto\delta_{IJ}\delta(t-t')$~\cite{parisi_sourlas,Nicolai:1980jc,Nicolai:1980js}. 

Now we may ask the question of how to interpret the case, when $F_I$ has non--zero curl and, therefore, cannot be expressed as $-\partial_I W$ (or, when it can, $W$ doesn't have continuous first derivatives). The answer is known: $F_I$ can be interpreted as a vector potential. 

More precisely, it's known, since  the work of Helmholtz, Clebsch and Monge, in the 19th century, that any vector field--under assumptions of smoothness, that ensure the existence of derivatives, of course--can be written as the sum of a curl--free part (that describes the effects of a scalar potential) and of a divergence--free part (that describes the effects of a vector potential)~\cite{lamb1993hydrodynamics}.

In ref.~\cite{Nicolis:2016osp} numerical evidence has been presented, that, when $F_I$ can be written as $-\partial_I W$, then the noise fields, in that case,  do, indeed, satisfy the identities that imply that the noise fields are Gaussian, with ultra--local 2--point function. In ref.~\cite{Nicolis:2017lqk} evidence was presented, that these identities also hold for target space, not only worldline, supersymmetry, when the superpotential can be globally defined. 

So the question then, quite naturally, arises, whether the corresponding `noise fields'' describe the consistent closure of the system, when $F_I$ admits the most general Helmholtz/Clebsch--Monge/Hodge decomposition and neither its divergence, nor its curl vanishes. In the following we shall discuss ``one particle mechanics'', where there is only one dependent variable and shall consider the case of an indefinite number of particles in a future work. 

In equations, the question is, whether the fluctuations transform the canonical partition function
\begin{equation}
\label{Zcanonical}
Z=\int\,[\mathscr{D}x_I]\,e^{-S}
\end{equation}
to
\begin{equation}
\label{Zlangevin}
Z_\mathrm{L}=\int\,[\mathscr{D}x_I]\,e^{-S}\,\left|\mathrm{det}\,\frac{\delta\eta^J}{\delta x_I}\right|
\end{equation}
which is equal to 1 (by a choice of units)--provided $Z_\mathrm{L}$ doesn't vanish and no boundaries appear in target space. 
This means that the zeromodes of the determinant--weighted by their relative phase!--remain of measure zero, also, when the $F_I$ vector field isn't a gradient and that the winding modes don't destabilize the system. 

Setting up the framework for using Monte Carlo simulations to test these statements, is the topic of the present paper. 

The simplest case to consider, is that of a uniform magnetic field, along the $z-$direction, when $F_1 = y=x_2$ and $F_2=-x=-x_1$. In that case, the ``potential energy'' comprises of two terms, a ``harmonic potential'',
\begin{equation}
\label{harmonicV}
V_2(x,y)=\frac{1}{2}\left(x^2+y^2\right)
\end{equation}
and the contribution from the angular momentum, 
\begin{equation}
\label{angularmomV}
V_L(x,y,\dot{x},\dot{y})=-\dot{x}y+x\dot{y}
\end{equation} 
that's not a total derivative (though it does contribute a constant to the action, for solutions of the equations of motion, since $dV_L/dt=0$, when $x(t)$ and $y(t)$ satisfy the classical equations of motion). 

However $V_L(x,y,\dot{x},\dot{y})$ can be, also, interpreted as the coupling of the velocity to external fields, namely a constant magnetic field. In this case only edge states correspond to unbounded motion. 

Of course, since the action is quadratic in $x(t)$ and $y(t)$, it is possible to compute the partition function exactly, as the determinant of the corresponding operator, which, since it is a non--zero constant, can be absorbed in the measure, as it is a constant. This is the way the system is ``closed''--and the choice of boundary conditions is crucial. 

This doesn't imply, however, that the correlation functions of the noise fields, $\eta_1=\dot{x}-y$ and $\eta_2=\dot{y}+x$, sampled with the Euclidian action, necessarily, satisfy the identities of Gaussian fields, with ultra--local propagator: $\langle\eta_I(t)\rangle=0$, $\langle\eta_I(t)\eta_J(t')\rangle=\delta_{IJ}\delta(t-t')$. 

In the present example this can be, however, consistently imposed, since it amounts to redefining the measure by a globally defined constant. 

In fact, what this example illustrates is that the equilibrium configuration, when $(x(0),y(0)\neq (0,0)$--which is the case when fluctutations are taken into account,  isn't a point, but the circle $x(t)^2+y(t)^2=x(0)^2+y(0)^2$, i.e. a (circular) string. 
{\em This} is the difference between the scalar and the vector potential: in the former case the equilibrium configurations are points, the minima of the scalar potential; in the latter case the equilibrium configurations are extended objects, since it isn't the scalar potential that determines them.

A more general example is provided by the Lorenz system
\begin{equation}
\label{lorenz}
\begin{array}{l}
\displaystyle
\frac{dx_1}{dt}=F_1=\sigma(x_2-x_1)\\
\displaystyle
\frac{dx_2}{dt}=F_2=x_1(r-x_3)-x_2\\
\displaystyle
\frac{dx_3}{dt}=F_3=-bx_3+x_1x_2
\end{array}
\end{equation}
It is well known that this system is ``dissipative'', since $\nabla\cdot\bm{F}=-\sigma-1-b < 0$--the constants $r,\sigma,b\geq 0$;  and $\bm{F}$ doesn't define a superpotential, since $\nabla\times\bm{F}\neq\bm{0}$. In fact it can be shown to describe the dynamics of a magnetic moment in a bath subject to an external torque (the so--called Landau--Lifshitz--Bloch--Bloembergen model) ~\cite{Tranchida:2015brk}. 
The non-trivial fact is that, here, as for the case of the constant magnetic field, the competition between the conservative forces and  the dissipation doesn't lead, in general, to point-like attractors in the space of states; these only exist for certain parameter values and there is a phase transition to a chaotic phase. In such a phase there are infinitely many periodic orbits and there have been attempts to use them to construct the measure they define, using perturbative field theoretic techniques~\cite{cvitanovic2000chaotic}. 

In this contribution we shall discuss another way to address this issue, that does not rely on perturbation theory.

To this end we shall follow the approach of refs.~\cite{Nicolis:2016osp,Nicolis:2017lqk}: We shall write a lattice action, in terms of the $\bm{x}$ and compute 
the identities that the correlation functions that the noise fields would be expected to satisfy, were the system consistently closed. 

The corresponding action for the particle, moving in three--dimensional space, is given by the expression
\begin{equation}
\label{Scont}
\begin{array}{l}
\displaystyle
S[x_I]=\int\,dt\,\left\{
\frac{1}{2}\left(\dot{x}_1-F_1\right)^2+\frac{1}{2}\left(\dot{x}_2-F_2\right)^2+\frac{1}{2}\left(\dot{x}_3-F_3\right)^2
\right\}=\\
\displaystyle
\int\,dt\,\left\{\frac{1}{2}\left(\dot{x}_1^2+\dot{x}_2^2+\dot{x}_3^2\right)+
\frac{1}{2}\left(F_1^2+F_2^2+F_3^2\right)-\left(\dot{x}_1 F_1+\dot{x}_2F_2+\dot{x}_3F_3\right)
\right\}=\\
\displaystyle
\int\,dt\,\left\{
-\frac{1}{2}\left(x_1\frac{d^2}{dt^2}x_1+x_2\frac{d^2}{dt^2}x_2+x_3\frac{d^2}{dt^2}x_3\right)+
\frac{1}{2}\left(F_1^2+F_2^2+F_3^2\right)-\left(\dot{x}_1 F_1+\dot{x}_2F_2+\dot{x}_3F_3 \right)+\right.\\
\displaystyle
\hskip1.6truecm \left.
\frac{d}{dt}\left[\frac{1}{2}\left(x_1\dot{x}_1+x_2\dot{x}_2+x_3\dot{x}_3\right)\right]
\right\}
\end{array}
\end{equation}
We recognize the (Euclidian) action of a particle in a potential, coupled to a ``magnetic field'', defined through the terms that are linear in the velocities and can't be identified as, exclusively, boundary terms. 

The scalar potential is bounded from below and confines at infinity, therefore the canonical partition function is well--defined (in fact since the Euclidian Lagrangian is a sum of non-negative terms, the lower bound is obvious; what's less obvious is that it confines at infinity, since the coupling to the magnetic field excites winding modes. This is what the stability of the Monte Carlo sampling checks.)

The noise fields we shall study are defined, in the continuum, by $\eta_I\equiv\dot{x}_I-F_I$ and shall try to check, whether the identities their correlation functions 
satisfy are affected as we cross the transition from the phase where the stable classical solutions are points to the phase where they are the strange attractor.

In ref.~\cite{Brink:1976uf} how to couple the spinning particle to an external electromagnetic field,  at the level of the classical action  as well as its canonical quantization, was discussed.  Translated to our case this result allows us to identify the expression of eq.~(\ref{Scont}) with the ``bosonic part'' of the action of a superparticle in a particular--external--electromagnetic field (cf. also~\cite{Bellucci:2011gn}). 

So what is of interest is to study how the fluctuations of the particle affect the supersymmetries of the classical case. Since the action is no longer quadratic in the fields, $x(t), y(t), z(t)$, we shall compute the correlation functions, that enter in the identities, using Monte Carlo simulations. To this end, we shall use a lattice discretization of the partition function. This is the topic of the following section. 

\section{The lattice action}\label{lattaction}
It's straightforward to write down the lattice action, 
\begin{equation}
\label{Slatt}
S_\mathrm{latt}[\varphi_{I,n}]=\sum_{n=0}^{N-1}\,\left\{
-\varphi_{I,n}\varphi_{I,n+1}+\varphi_{I,n}^2+\frac{1}{2}F_{I,n}^2-(\varphi_{I,n+1}-\varphi_{I,n})F_{I,n}
\right\}
\end{equation}
that corresponds to the discretization of the expression in eq.~(\ref{Scont}), with $\varphi_{I,n}=a^{-1/2}x_{I,n}$ and the lattice spacing, $a$, has been absorbed in the other terms. The sum over $I=1,2,3$ is implicit.   The manipulations that led to eq.~(\ref{Scont}) simply had as purpose to produce a canonical kinetic term, a sensible potential energy and a coupling to well--defined sources. The latter are the new features, in fact. The action describes a particle in a non--uniform, external,  magnetic field. 

It should be stressed that the last term of the lattice action, when periodic boundary conditions are imposed, is sensitive only to lattice artifacts--terms that are proportional to positive powers of the lattice spacing--and contributions that are not total derivatives, in the continuum limit. It is the latter that are of interest in the present case, of course. 

The question, that remains to be addressed, is, whether, upon coupling this system to a bath, the system, along with the bath, this time,  remains closed--which, in turn, implies that the bath does describe {\em all} the fluctuations. 

To address this question, we shall use the lattice action to compute the correlation functions of the noise fields and check the identities they would satisfy, were the system, indeed, closed. 

The noise fields, on the lattice,  are given by the expressions
\begin{equation}
\label{noisefields}
\begin{array}{l}
\displaystyle
\eta_{1,n} = \varphi_{1,n+1}-\varphi_{1,n}-F_{1,n}\\
\displaystyle
\eta_{2,n} =\varphi_{2,n+1}-\varphi_{2,n}-F_{2,n}\\
\displaystyle
\eta_{3,n}=\varphi_{3,n+1}-\varphi_{3,n}-F_{3,n}\\
\end{array}
\end{equation}
The identities to be checked are:
\begin{equation}
\label{identities}
\begin{array}{l}
\left\langle\eta_{I,n}\right\rangle=\left\langle F_{I,n}\right\rangle\stackrel{?}{=}0\\
\left\langle(\eta_{I,n}-\left\langle\eta_{I,n}\right\rangle)(\eta_{J,n'}-\left\langle\eta_{J,n'}\right\rangle)\right\rangle\stackrel{?}{\propto}\delta_{IJ}\delta_{n,n'}\\
\left\langle\left(\eta_{I,n}-\left\langle\eta_{I,n}\right\rangle\right)^4\right\rangle\stackrel{?}{=}3\left\langle\left(\eta_{I,n}-\left\langle\eta_{I,n}\right\rangle\right)^2\right\rangle^2
\end{array}
\end{equation}
Now it is clear that $\langle\eta_{I,n}\rangle=0$, can hold, in the non--chaotic, as well as in the chaotic, phases.  It should be stressed that $\langle\varphi_{I,n}\rangle\neq 0$, however. What is important to understand is that, while the expectation values are defined with respect to $e^{-S_\mathrm{latt}}$, the fluctuations do lead to the full measure, that will be the chaotic attractor in the corresponding phase. 

These define, in terms of the ``physical fields'', $x_I$, deduced from the $\varphi_{I,n}$, in the scaling limit, new identities between the correlation functions on the attractor in the chaotic phase. In that case they imply that supersymmetry, according to the usual criterion, is not broken. The subtleties pertaining to its possible breaking are encoded in the possible ``anomalies'' of the multi--point functions of the noise fields, that deserve a detailed study, that, while time-consuming, doesn't entail any conceptual issues, since the expectation values are computed using the canonical partition function of the Euclidian action, that's bounded from below and confines at infinity--and the periodic boundary conditions ensure that the edge states don't destablilize the system.

\section{Conclusions}\label{concl}
In the present paper we have described the framework for studying how  ordinary differential equations, whose RHS isn't a gradient and, therefore, describe the motion of a particle in the presence of fluxes, can be consistently considered as  defining backgrounds for the dynamics of physical systems, whose properties can be described by supersymmetric theories, that provide the consistent closure for the fluctuations. This is worldline supersymmetry;  the target space geometry is that of a $d=1$ supersymmetric non--linear $\sigma-$model and it will be interesting to explore how the techniques presented, for instance, in refs.~\cite{Delduc:2019rpd,Ivanov:2013fha,Lyakhovich:1996we} can be applied, in order to deduce how target space supersymmetry might hold. 

The new feature, in the presence of fluxes, is that  the equilibrium configurations, in target space  need not be  points, but can be extended objects, namely, strings. Whereas much of the focus of chaotic dynamical systems has been on the attractor, as the solution of the equations of motion, what are its properties in the presence of fluctuations, i.e. beyond the equations of motion, is defined by the correlation functions of the noise fields, defined in terms of the ``physical'' fields, that appear in the differential equations. This is, of course, but the first step for describing the fluctuations of systems, whose classical limit shows deterministic chaos (cf. also~\cite{ovchinnikov2016introduction}). 
 
 This approach extends the insights of~\cite{Fayet:1975ki,ORaifeartaigh:1975nky}, who noted that, with, at least,  three superfields, supersymmetry is, generically, spontaneously broken, in that the 1--point functions of the auxiliary fields cannot all be made to vanish. 
 For the Lorenz attractor we understand that the 1--point functions, $\langle\eta_{I,n}\rangle=\langle F_{I,n}\rangle$ can vanish for all attractors, chaotic or non--chaotic; so the properties of the multi--point functions of the noise fields are, in fact,  relevant for describing how supersymmetry is realized--or broken. The details of this investigation will be presented in forthcoming work.

 {\bf Acknowledgements:} It's a pleasure to thank the organizers of the workshop ``Supersymmetries and Quantum Symmetries 2019'' in Yerevan for the warm hospitality and the stimulating discussions.

\bibliographystyle{utphys}
\bibliography{SUSY}

\providecommand{\href}[2]{#2}\begingroup\raggedright\begin{thebibliography}{10}

\bibitem{parisi_sourlas}
G.~Parisi and N.~Sourlas, ``{Supersymmetric Field Theories and Stochastic
  Differential Equations},''
\href{http://dx.doi.org/10.1016/0550-3213(82)90538-7}{{\em Nucl. Phys.}
  {\bfseries B206} (1982) 321--332}.

\bibitem{Nicolai:1980jc}
H.~Nicolai, ``{Supersymmetry without anticommuting variables},'' in {\em
  {Unification of the fundamental particle interactions. Proceedings,
  Europhysics study conference, Erice, Italy, March 17-24, 1980}}, p.~689.
\newblock
1980.
\newblock

\bibitem{Nicolai:1980js}
H.~Nicolai, ``{Supersymmetry and Functional Integration Measures},''
\href{http://dx.doi.org/10.1016/0550-3213(80)90460-5}{{\em Nucl. Phys.}
  {\bfseries B176} (1980) 419--428}.

\bibitem{lamb1993hydrodynamics}
H.~Lamb, {\em Hydrodynamics}.
\newblock Cambridge University Press, 1993.

\bibitem{Nicolis:2016osp}
S.~Nicolis, ``{How quantum mechanics probes superspace},''
  \href{http://arxiv.org/abs/1606.08284}{{\ttfamily arXiv:1606.08284
  [hep-th]}}.
[Phys. Part. Nucl. Lett.14,no.2,357(2017)].

\bibitem{Nicolis:2017lqk}
S.~Nicolis, ``{Probing the holomorphic anomaly of the $D=2, \mathcal{N}=2$,
  Wess-Zumino model on the lattice},''
  \href{http://dx.doi.org/10.1134/S1063779618050313}{{\em Phys. Part. Nucl.}
  {\bfseries 49} no.~5, (2018) 899--903},
\href{http://arxiv.org/abs/1712.07045}{{\ttfamily arXiv:1712.07045 [hep-th]}}.

\bibitem{Tranchida:2015brk}
J.~Tranchida, P.~Thibaudeau, and S.~Nicolis, ``{Quantum Magnets and Matrix
  Lorenz Systems},''
  \href{http://dx.doi.org/10.1088/1742-6596/574/1/012146}{{\em J. Phys. Conf.
  Ser.} {\bfseries 574} (2015) 012146},
\href{http://arxiv.org/abs/1504.06161}{{\ttfamily arXiv:1504.06161 [math.MP]}}.

\bibitem{cvitanovic2000chaotic}
P.~Cvitanovi{\'c}, ``Chaotic field theory: a sketch,'' {\em Physica A:
  Statistical Mechanics and its Applications} {\bfseries 288} no.~1-4, (2000)
  61--80, \href{http://arxiv.org/abs/nlin/0001034}{{\ttfamily
  arXiv:nlin/0001034 [nlin.CD]}}.

\bibitem{Brink:1976uf}
L.~Brink, P.~Di~Vecchia, and P.~S. Howe, ``{A Lagrangian Formulation of the
  Classical and Quantum Dynamics of Spinning Particles},''
\href{http://dx.doi.org/10.1016/0550-3213(77)90364-9}{{\em Nucl. Phys.}
  {\bfseries B118} (1977) 76--94}.

\bibitem{Bellucci:2011gn}
S.~Bellucci, N.~Kozyrev, S.~Krivonos, and A.~Sutulin, ``{${\cal N}=4$ Chiral
  Supermultiplet Interacting with A Magnetic Field},''
  \href{http://dx.doi.org/10.1103/PhysRevD.85.065024}{{\em Phys. Rev.}
  {\bfseries D85} (2012) 065024},
\href{http://arxiv.org/abs/1112.0763}{{\ttfamily arXiv:1112.0763 [hep-th]}}.

\bibitem{Delduc:2019rpd}
F.~Delduc and E.~Ivanov, ``{${\cal N}{=}4$ supersymmetric $d=1$ sigma models on
  group manifolds},''
  \href{http://dx.doi.org/10.1016/j.nuclphysb.2019.114806}{{\em Nucl. Phys.}
  {\bfseries B949} (2019) 114806},
\href{http://arxiv.org/abs/1907.09518}{{\ttfamily arXiv:1907.09518 [hep-th]}}.

\bibitem{Ivanov:2013fha}
E.~A. Ivanov and A.~V. Smilga, ``{Quasicomplex ${\cal N}=2, d=1$ Supersymmetric
  Sigma Models},'' \href{http://dx.doi.org/10.3842/SIGMA.2013.069}{{\em SIGMA}
  {\bfseries 9} (2013) 069},
\href{http://arxiv.org/abs/1302.2902}{{\ttfamily arXiv:1302.2902 [hep-th]}}.

\bibitem{Lyakhovich:1996we}
S.~L. Lyakhovich, A.~{\relax Yu}. Segal, and A.~A. Sharapov, ``{A Universal
  model of $D = 4$ spinning particle},''
  \href{http://dx.doi.org/10.1103/PhysRevD.54.5223}{{\em Phys. Rev.} {\bfseries
  D54} (1996) 5223--5238},
\href{http://arxiv.org/abs/hep-th/9603174}{{\ttfamily arXiv:hep-th/9603174
  [hep-th]}}.

\bibitem{ovchinnikov2016introduction}
I.~V. Ovchinnikov, ``Introduction to supersymmetric theory of stochastics,''
  {\em Entropy} {\bfseries 18} no.~4, (2016) 108.

\bibitem{Fayet:1975ki}
P.~Fayet, ``{Spontaneous Supersymmetry Breaking Without Gauge Invariance},''
\href{http://dx.doi.org/10.1016/0370-2693(75)90730-3}{{\em Phys. Lett.}
  {\bfseries 58B} (1975) 67}.

\bibitem{ORaifeartaigh:1975nky}
L.~O'Raifeartaigh, ``{Spontaneous Symmetry Breaking for Chiral Scalar
  Superfields},''
\href{http://dx.doi.org/10.1016/0550-3213(75)90585-4}{{\em Nucl. Phys.}
  {\bfseries B96} (1975) 331--352}.

\end{thebibliography}\endgroup
\end{document}